\UseRawInputEncoding
\documentclass[twocolumn]{aastex701}
\usepackage{longtable}
\usepackage[flushleft]{threeparttable}
\usepackage{tabularx}
\usepackage{multirow}
\usepackage{graphicx}
\usepackage{amsmath,amssymb}
\usepackage{color}
\usepackage{units}
\usepackage{epstopdf}
\usepackage{hyperref}
\usepackage{multirow}
\usepackage{url}
\usepackage{footnote}
\usepackage{enumitem}\setlist[description]{font=\textendash\enskip\scshape\bfseries}

\newcommand{\beq}{\begin{equation}}
\newcommand{\eeq}{\end{equation}}
\newcommand{\bdm}{\begin{displaymath}}
\newcommand{\edm}{\end{displaymath}}
\newcommand{\T}[1]{\tilde{#1}}

\definecolor{Gray}{gray}{0.9}
\definecolor{orange}{rgb}{0.9,0.5,0}

\newcommand{\ztfink}[1]{ZTF-Fink}

\begin{document}
\title{The ZTF-ULTRASAT experiment: Characterizing the non-transients in ULTRASAT's high cadence survey}

\author[orcid=0009-0000-0467-6911]{Daniel Warshofsky}
\email[show]{warsh029@umn.edu}
\affiliation{School of Physics and Astronomy, University of Minnesota, Minneapolis, MN 55414}
\author[orcid=0000-0002-8262-2924]{Michael W. Coughlin}
\email[show]{cough052@umn.edu}
\affiliation{School of Physics and Astronomy, University of Minnesota, Minneapolis, MN 55414}
\author[orcid=0009-0003-6181-4526]{Theophile Jegou Du Laz}
\email{tdulaz@caltech.edu}
\affiliation{Division of Physics, Mathematics, and Astronomy, California Institute of Technology, Pasadena, CA 91125, USA}
\author[orcid=0000-0002-9017-3567]{Anna Y. Q. Ho}
\email{ayh24@cornell.edu}
\affiliation{Cornell University}
\author[orcid=0000-0003-1673-970X]{S. Bradley Cenko}
\email{brad.cenko@nasa.gov}
\affiliation{Astrophysics Science Division, NASA Goddard Space Flight Center, Greenbelt, MD 20771, USA}
\affiliation{Joint Space-Science Institute, University of Maryland, College Park, MD 20742, USA}
\author{Andrew Drake}
\email{ajd@astro.caltech.edu}
\affiliation{Division of Physics, Mathematics, and Astronomy, California Institute of Technology, Pasadena, CA 91125, USA}
\author[orcid=0000-0003-1546-6615]{Jesper Sollerman}
\email{jesper@astro.su.se}
\affiliation{The Oskar Klein Centre, Department of Astronomy, Stockholm University SE-106 91 Stockholm, Sweden}
\author[orcid=0000-0001-7357-0889]{Argyro Sasli}
\email{asasli@umn.edu}
\affiliation{School of Physics and Astronomy, University of Minnesota, Minneapolis, MN 55414}
\affiliation{NSF Institute on Accelerated AI Algorithms for Data-Driven Discovery (A3D3)}
\author[orcid=0000-0001-7648-4142]{Ben Rusholme}
\email{rusholme@ipac.caltech.edu}
\affiliation{IPAC, California Institute of Technology, 1200 E. California Blvd, Pasadena, CA 91125, USA}
\author[orcid=0000-0002-8532-9395]{Frank J. Masci}
\email{fmasci@ipac.caltech.edu}
\affiliation{IPAC, California Institute of Technology, 1200 E. California Blvd, Pasadena, CA 91125, USA}
\author[orcid=0000-0001-7062-9726]{Roger Smith}
\email{rsmith@astro.caltech.edu}
\affiliation{Caltech Optical Observatories}
\author[orcid=0000-0001-7681-4316]{A.M. Krassilchtchikov}
\email{aleksandr.krasilshchikov@weizmann.ac.il}
\affiliation{Department of Particle Physics and Astrophysics, Weizmann Institute of Science, Rehovot 7610001, Israel}
\author[orcid=0000-0002-2918-1824]{David Berge}
\email{david.berge@desy.de}
\affiliation{Deutsches Elektronen-Synchrotron DESY, Humboldt-University Berlin}
\author[orcid=0000-0002-6786-8774]{Eran O. Ofek}
\email{eran.ofek@weizmann.ac.il}
\affiliation{Department of Particle Physics and Astrophysics, Weizmann Institute of Science, Rehovot 7610001, Israel}
\author[orcid=0000-0003-1525-5041]{Yossi Shvartzvald}
\email{yossi.shvartzvald@weizmann.ac.il}
\affiliation{Department of Particle Physics and Astrophysics, Weizmann Institute of Science, Rehovot 7610001, Israel}
\author[orcid=0000-0002-0387-370X]{Reed L. Riddle}
\email[0000-0002-0387-370X]{riddle@caltech.edu}
\affiliation{Division of Physics, Mathematics, and Astronomy, California Institute of Technology, Pasadena, CA 91125, USA}
\author[orcid=0000-0002-5619-4938]{Mansi M. Kasliwal}
\affiliation{Division of Physics, Mathematics, and Astronomy, California Institute of Technology, Pasadena, CA 91125, USA}
\email{mansi@astro.caltech.edu}
\author[0000-0002-3168-0139]{Matthew J. Graham}
\affiliation{California Institute of Technology}
\email{mjg@caltech.edu}
\author[0000-0001-8018-5348]{Eric C. Bellm}
\affiliation{DIRAC Institute, Department of Astronomy, University of Washington, 3910 15th Avenue NE, Seattle, WA 98195, USA}
\email{ecbellm@uw.edu}
\begin{abstract}
The forthcoming launch of the Ultraviolet Transient Astronomy Satellite (ULTRASAT) will transform our understanding of the transient ultraviolet sky by increasing our ability to identify  transients due to its unprecedented 204 deg$^{2}$ field of view. While rapid (extragalactic) transients are a priority science area for the mission, flaring stars and AGN can often contaminate searches for such objects. To prepare for these challenges, the Zwicky Transient Facility (ZTF)-ULTRASAT experiment observed five fields at high cadence over three nights, in close proximity to ULTRASAT’s three northern high-cadence fields. A real-time filter identified seven transient candidates, of which five were persistent variable sources and two were spurious. Periods and amplitudes derived from the ZTF Source Classification Project (SCoPe) showed that three candidates were RR Lyrae stars with short periods and high amplitudes, while the remaining two displayed flaring behavior. We demonstrate that short-timescale, high-amplitude variables can systematically mimic transient alerts in high-cadence UV surveys, and we provide a concrete strategy to mitigate this contamination using pre-existing machine learning catalogs. 
\end{abstract}



\section{Introduction}

ULTRASAT, the Ultraviolet Transient Astronomy Satellite \citep{Shvartzvald24}, is an upcoming space-based ultraviolet observatory. With a 204 deg$^{2}$ field of view (FOV), ULTRASAT is poised to expand our understanding of the time-domain sky in the ultraviolet (UV). Its large FOV will enable the discovery of numerous transients—astrophysical sources that appear and fade on timescales of human lifetimes or shorter. ULTRASAT’s UV sensitivity will allow the detection of phenomena such as kilonovae \citep{Shvartzvald24}, shock breakouts in early SNII light curves \citep{UltrasatSN2_1,UltrasatSN2_2}, and flares from M-dwarfs \citep{rekhi2025censusnuvmdwarfflares} at rates far exceeding previous capabilities. However, false alerts can arise from hardware or software issues, including cosmic rays, incorrect difference images, or suboptimal astrometry, while non-transient astrophysical sources such as asteroids or previously unknown variable stars can also trigger alerts.

To anticipate the data volume and associated false alarms, we conducted a three-night experiment with the Zwicky Transient Facility (ZTF) \citep{ZTFObservingSystem,ZTFSystemOverview} from June 04 2024 to June 07 2024, designed to emulate some of the challenges ULTRASAT will face. ZTF is a ground-based time-domain survey that has discovered numerous transients and variable sources \citep{2023arXiv230707618R,10.1093/mnras/stae1164,Chen_2020}. One of the main contaminants for transient searches are periodic sources, particularly those with short periods and high amplitudes. Although ZTF’s FOV is smaller than ULTRASAT’s (~47 deg$^{2}$. vs. ~204 deg$^{2}$.), their nominal limiting magnitudes are similar (Table~\ref{table:ZTFvsULTRASAT}), and in this experiment, ZTF’s cadence was adjusted to match ULTRASAT’s high-cadence fields. While ULTRASAT’s UV coverage does not overlap with ZTF’s optical filters (Fig.~\ref{fig:ZTF_filters}) and no large ground-based UV surveys exist, ZTF remains a valuable proxy for exploring ULTRASAT’s data processing challenges.

Optical surveys like ZTF produce alerts at rates too high for manual inspection. Alert brokers are therefore used to efficiently distribute and filter these data. Alert packets, often simply called alerts, are distributed to the broader community. ZTF employs the Python-based Kowalski\footnote{https://github.com/skyportal/Kowalski} alert broker and BOOM, a new Rust-based broker \citep{Boom}. Even with these systems, individual groups must filter alerts to identify sources of interest. Traditional methods such as color thresholds or cross-matching external catalogs are often insufficient for robust source classification, making machine learning an essential tool. ZTF has developed several machine-learning frameworks, including the Bright Transient Survey (BTS) Bot \citep{BTSBot}, the Source Classification Project (SCoPe) \citep{SCoPeI}, and APPLECIDER \citep{CiderI}. BTS-Bot focuses on rapid transient classification, SCoPe characterizes variable source properties, and APPLECIDER develops foundational models for broader astrophysical applications. For the ZTF-ULTRASAT experiment, SCoPe \citep{SCoPeIII} was used to analyze the variable properties of false positives. 

The structure of the paper is as follows. Sec.~\ref{sec:SCoPe} provides a brief overview of the SCoPe project and its classification scheme, while Secs.~\ref{experiment} and \ref{sec:science} describe the ZTF-ULTRASAT experiment and review the candidates identified during the study.

\begin{table}
    \centering
    \caption{ZTF vs ULTRASAT capabilities}
    \begin{tabular}{ |>{\centering\arraybackslash}m{5em} | >{\centering\arraybackslash}m{7em} |>{\centering\arraybackslash}m{7em} | } 
      \hline
      & ZTF (ZTFg)& ULTRASAT\textsuperscript{a}\\ 
      \hline
      FOV (sqr. deg) & 47.7 &  204\\ 
      \hline
      Slew Speed & 50 deg/min&  $>$ 30 deg/min\\
      \hline
      Pixel scale & 1\farcs01 pixel$^{-1}$ &  5\farcs4 pixel$^{-1}$\\
      \hline
      5$\sigma$ Limiting Magnitude & 20.8 30s exp &  22.5 900s (AB 20,000K BB) \\
      \hline
      Wavelength Coverage &3670--5610 \AA &  2300-2900 \AA \\ 
      \hline
    \end{tabular}
    \\
    \vspace{1mm} 
    \textsuperscript{a}For the central 170 deg$^{2}$.
    \label{table:ZTFvsULTRASAT}
\end{table}

\section{SCoPe}
\label{sec:SCoPe}

\texttt{SCoPe} is the current effort to create a catalog of the properties of all variable sources in the northern sky observable by ZTF using machine learning \citep{SCoPeI, SCoPeII, SCoPeIII}. SCoPe produces 44 unique classifications for each light curve, which are grouped into two main types: phenomenological and ontological. Ontological classifications correspond to known astrophysical source types, such as active galactic nuclei (AGN), Cepheids, or Cataclysmic Variables (CVs), while phenomenological classifications describe abstract light-curve properties, such as flaring, periodicity, or eclipsing behavior. Each classification is computed by two machine learning algorithms: a deep neural network (DNN) \citep{DNN} and XGBoost (XGB) gradient-boosted decision trees \citep{XGB}.

Light-curve data from ZTF, like that from all ground-based optical surveys, is non-uniform: the number of observations and time coverage varies across sources. Because SCoPe's machine learning models cannot directly handle this non-uniformity, it computes a fixed set of features for each light curve \citep{SCoPeII}. ZTF releases data in batches, called data releases, and SCoPe uses only the most recent stable release at training time—DR16 in this case—to avoid out-of-distribution effects. Features are divided into three categories: simple statistical summaries of the light curve (e.g., mean, variance), auxiliary metadata from ZTF and other surveys, and the period of the light curve.

\texttt{SCoPe} employs three period-finding algorithms -- Lomb-Scargle (LS) \citep{Lomb,Scargle}, Conditional Entropy (CE) \citep{CE}, and Analysis of Variance (AOV) \citep{AOV} -- to capture different types of periodic variability. Periods are computed on a uniform linear frequency grid from 1/1800 Hz\footnote{set for computational reasons} (30 minute period) to the Nyquist frequency (half the baseline of the light curve), with common aliases at one day, one month, and one year removed. The most significant periods, their associated significances, and tie-broken top periods from LS and CE are used as features. Additionally, the DNN models use a dmdt feature \citep{dmdt}, a 2D histogram binning all pairs of light-curve points by baseline and magnitude difference. This transforms the light curve into an image suitable for convolutional layers.

DNNs consist of interconnected layers performing linear transformations followed by nonlinear activation functions. SCoPe’s DNN models have two branches: one with fully connected dense layers and dropout layers to prevent overfitting, and another with convolutional layers that process the dmdt histogram. These branches are combined through additional dense layers with dropout, and final outputs are clamped between 0 and 1 via a sigmoid activation. In contrast, XGB leverages decision trees for classification. Trees partition the feature space to separate positive and negative examples, and XGB iteratively improves classification by computing the gradient of the loss function with respect to the current tree ensemble. This yields high performance but increases susceptibility to overfitting. Unlike DNNs, XGB provides a more direct measure of feature importance for each classification. Performance of \texttt{SCoPe}'s algorithms can be seen in figures 7,8 and 9 of \cite{SCoPeIII}. 

\begin{figure}[!htb]
    \centering
    \includegraphics[width=1\linewidth]{{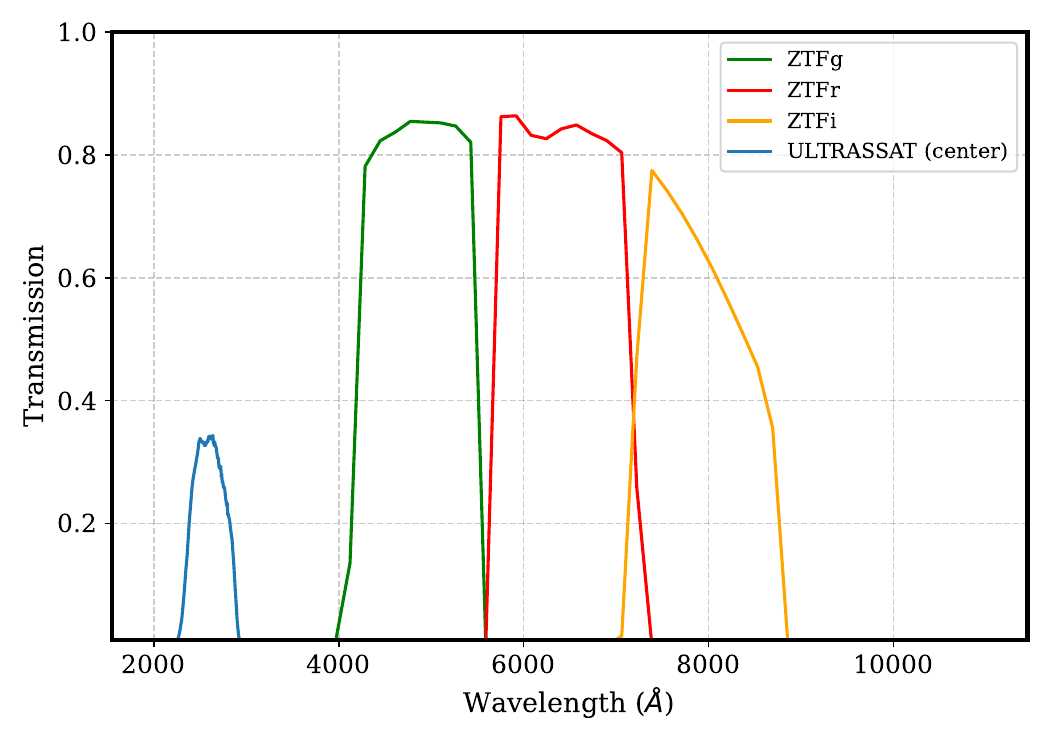}}
    \caption{The depicted trasmission for ULTRASAT is the full throughput of the entire optical system. ULTRASAT will observe wavelengths that are impossible to observe with ZTF.}
    \label{fig:ZTF_filters}
\end{figure}

Classifications are computed per light curve, not per astrophysical source. ZTF uses three filters, ZTFg, ZTFr, and ZTFi (Fig.~\ref{fig:ZTF_filters}) and \texttt{SCoPe} treats each filter separately, so a single source may have up to three classified light curves. PSF photometry may further split sources, particularly blended sources, across different light curves even within the same filter. Additionally, ZTF tiles the sky using two overlapping static grids (primary and secondary), meaning many sources have multiple light curves. As a rule of thumb, one million light curves classified as periodic likely correspond to $\sim$500,000 astrophysical sources.

Training used 170,632 manually curated light curves \citep{SCoPeI}, with an iterative process of expert review through a graphical interface \citep{WaCr2019,SkyPortalCoughlin_2023} to refine the dataset. Classifiers were trained for labels with at least 50 positive examples using an 81-9-10 training-validation-test split. Both DNN and XGB models perform comparably for well-represented classes. For each light curve, \texttt{SCoPe} outputs a score between 0 and 1 for each classification, with 1 indicating maximum confidence. Since all classifiers are independent, a light curve receives 88 scores (44 DNN + 44 XGB), which do not need to sum to 1. This allows non-mutually exclusive classifications, enabling hierarchical labeling: for instance, an RR Lyrae C star is simultaneously labeled RR Lyrae, pulsator, periodic, and variable. The full classification trees are shown in Figs. 2 and 3 of \cite{SCoPeIII}.

\section{ZTF-ULTRASAT experiment}
\label{experiment}
\begin{figure}[!htb]
    \centering
    \includegraphics[width=1\linewidth]{{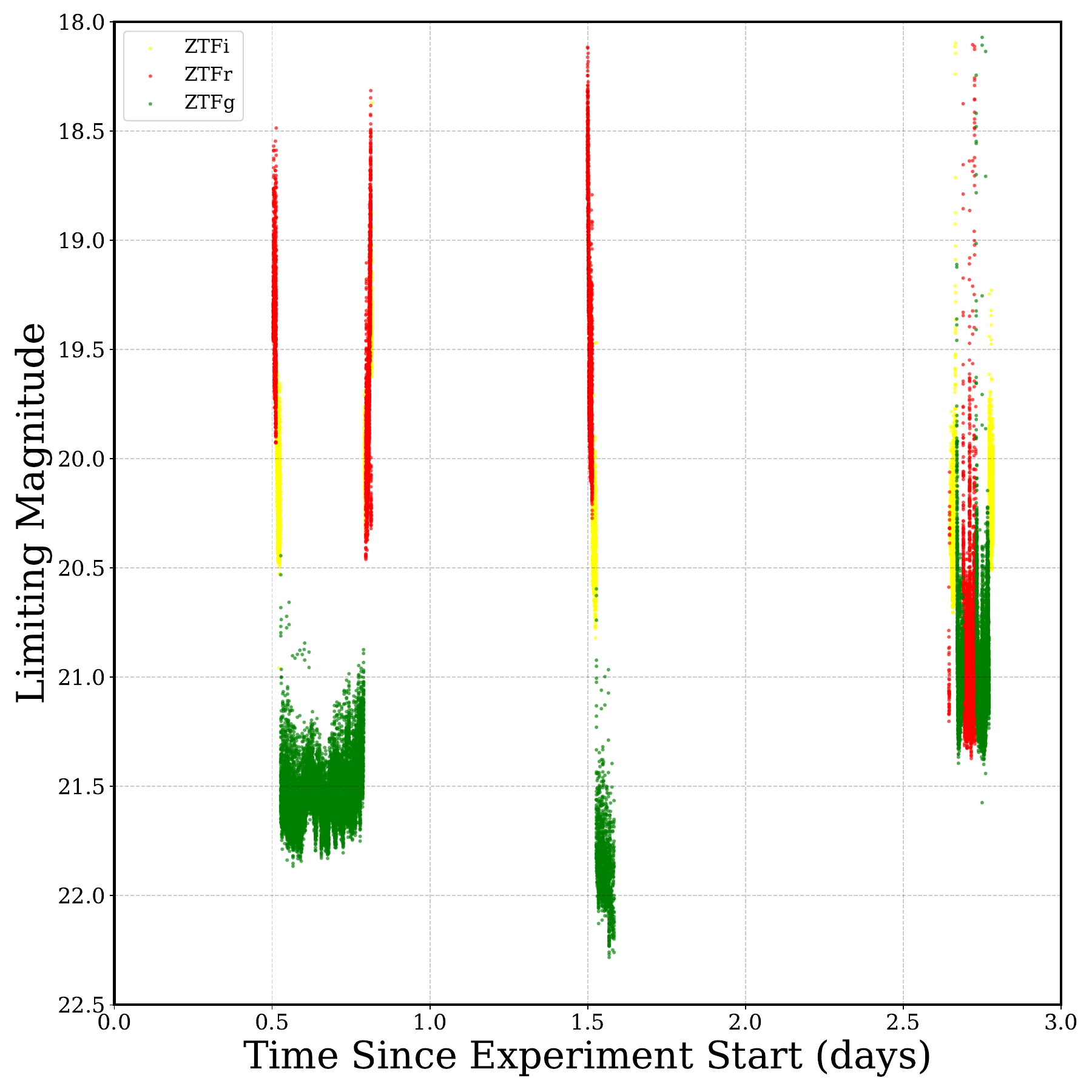}}
    \caption{5$\sigma$ Limiting magnitude of all the exposures captured during the experiment duration. A limiting magnitude of ~20.5 is typical for ZTF's normal survey \citep{Graham_2019}.}
    \label{fig:LM_overtime}
\end{figure}
\begin{figure}[!htb]
    \centering
    \includegraphics[width=1\linewidth]{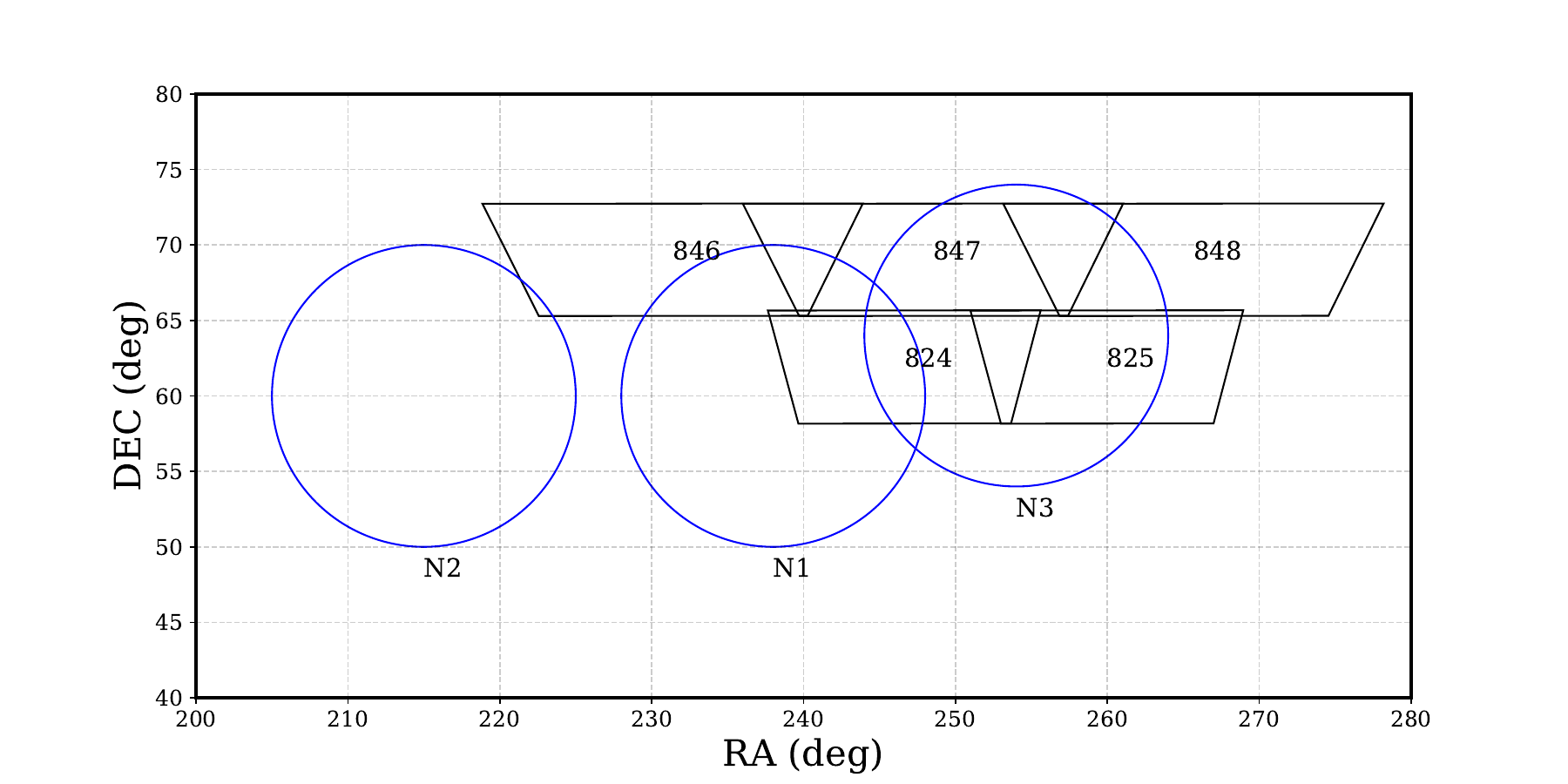}
    \caption{ULTRASAT's north high cadence fields N1,N2 and N3 as blue circles and the ZTF-ULTRASAT experiment fields as the boxes. The ZTF fields are labeled with their field id number. Note that while the N1 and N3 fields overlap with the experiment, N2 does not have overlap with the experiment fields.}
    \label{fig:ULTRASAT_FIELDS}
\end{figure}

The ZTF-ULTRASAT experiment was conducted June 4th 2024 at 04:00 UTC to June 7th 2024 at 10:15 UTC. The limiting magnitude during the experiment can be seen in Fig.~\ref{fig:LM_overtime}. The experiment consisted of observing ZTF fields 825, 824, 847, 848 and 846, which overlap with the ULTRASAT high cadence fields N1 and N3 which correspond to $\sim$\,238$^\circ$ RA, $\sim$\,60$^\circ$ Dec and $\sim$\,254$^\circ$ RA, $\sim$\,64$^\circ$ Dec respectively (Fig.~\ref{fig:ULTRASAT_FIELDS}).The N3 field will be observed by ULTRASAT in its first year and N2 in the second year. N1 will be considered for observation in ULTRASAT's third year or beyond\footnote{https://www.weizmann.ac.il/ultrasat/science-mission/modes-of-operation/modes-of-operation
}. The exposure times for the ZTF-ULTRASAT experiment were  60\,s, 120\,s and 180\,s for each night respectively, with a goal of measuring transients as they potentially faded over the course of the experiment. 

ZTF finds transients by performing difference imaging relative to a reference catalog, which is formed from a stack of earlier, high quality ZTF images \citep{masci,image_sub}. When an image is taken, a difference image is made from its corresponding reference. Positive residuals in the difference images with significance greater than 5$\sigma$ generate alerts \citep{patterson}. An alert is a packet of information that contains image cut outs of the area of interest with position and photometry information that is pushed through the alert broker \texttt{Kowalski} \citep{kowalski}. The alert broker ships the information to the wider community and allows for basic filtering on provided meta data. 

The main pathway ZTF visualizes these alerts is with \texttt{Fritz}, an instance of the \texttt{SkyPortal} platform \citep{vanderWalt2019,SkyPortalCoughlin_2023}, which is a web based platform that has a graphical user interface for accessing  photometry, alerts and spectra while also providing tools to perform streamlined manual scanning, basic analysis and trigger followup on interesting sources.  During the experiment, any alerts generated were passed through a filter specifically designed for the experiment.

The filter included two criteria, i.) a counterpart within 1.5$"$ classified as a star OR within 5$"$ classified as a galaxy and ii.) an alert that is $\gtrsim$ 0.8\,mag brighter than a previous detection in the same filter, where ``previous'' means 0.015 -- 1 day prior, and the previous detection is brighter than the closest source in the reference catalog. Both conditions need to be met to pass the filter. If the filter was passed, the alerts were put in a database and manually scanned through \texttt{Fritz} to remove obvious outliers or asteroids. The remaining sources were gathered and analyzed for any periodic properties using a modified version of the \texttt{SCoPe} pipeline. 

A modified version of the \texttt{SCoPe} \citep{SCoPeII} feature generation pipeline was used to compute periods for all sources in the experiment fields. For the experiment, sources that exhibit a significant change in magnitude ($\gtrsim$ 5-$\sigma$) are pushed to the public as a ZTF alert and added to the ZTF alert catalog. Our pipeline used all of the alerts generated from the experiment area during the experiment in addition to data from cross matching external catalogs like GAIA \citep{gaia1,gaia2},, ALLWISE \citep{wise,neowise} and PS1 \citep{PS1}. Once all of the data was aggregated, only period finding was performed. Typically, the high cadence observations, e.g. observations within 30 minutes of each other, are removed from light curves for analysis of their periodicity, but since all observations were performed at similarly high cadences for the ZTF-ULTRASAT experiment, this step was not performed. Additionally, the frequency grid was expanded to include periods as low as 5 minutes. The \texttt{SCoPe} classifiers are only trained on archival data through DR16, so they cannot be used to directly classify light curves from the alert stream. The candidates from the experiment were spatially cross-matched with the existing \texttt{SCoPe} catalog for classifications.

\section{Candidates}
\label{sec:science}

We now provide a description of the 7 candidates that passed our filter during the experiment. A summary of the candidates can be found in Table \ref{table:Candidates}.

\begin{table*}
\centering
    \caption{Transient Candidates}
\begin{tabular}{|>{\centering\arraybackslash}m{7em}|>{\centering\arraybackslash}m{6em}|>{\centering\arraybackslash}m{6em}|>{\centering\arraybackslash}m{7em}|>{\centering\arraybackslash}m{7em}|>{\centering\arraybackslash}m{3em}|>{\centering\arraybackslash}m{3em}|>{\centering\arraybackslash}m{6em}|}
\hline
Candidates   & Period SCoPe (days) & Period Experiment (days) &SCoPe Classification XGB &SCoPe Classification DNN &Ra (deg)&Dec (deg)& External Reference \\
\hline
ZTF24aaqqtht & Not found                  & Not found                       & None                   & None                   &255.45&63.38&  NA                   \\
\hline
ZTF24aaqrjdd &Not found                  & Not found                       & None                   & None                   &238.05&72.10&   NA                 \\
\hline
ZTF18aajtlgu & 1.533               & 0.255                    & RR Lyrae                       & RR Lyrae                       &264.39&61.766& \cite{Sesar_2017}        \\
\hline
ZTF18aajtkma & Not found   & Not found        & Flaring,                      & CV                           &262.53&62.79& \cite{aajkma_0riginal}       \\
\hline
ZTF18aakfqxu & 0.483               & 0.483                    & RR Lyrae AB                    & RR Lyrae AB                    &226.20&66.27& \cite{Sesar_2013}        \\
\hline
ZTF18aapnpxp & 1.471               & 0.163                    & RR Lyrae                       & RR Lyrae                       &227.46&67.23& \cite{aapnpxp_original}         \\
\hline
ZTF21aasjkbd & 1.216                 & Not found                       & Irregular                            & Flaring                            &243.87& 59.23& NA       
\\
\hline
    \end{tabular}
    \label{table:Candidates}
\end{table*}

\subsection{ZTF24aaqqtht}
\begin{figure}[!htb]
    \centering
    \includegraphics[width=1\linewidth]{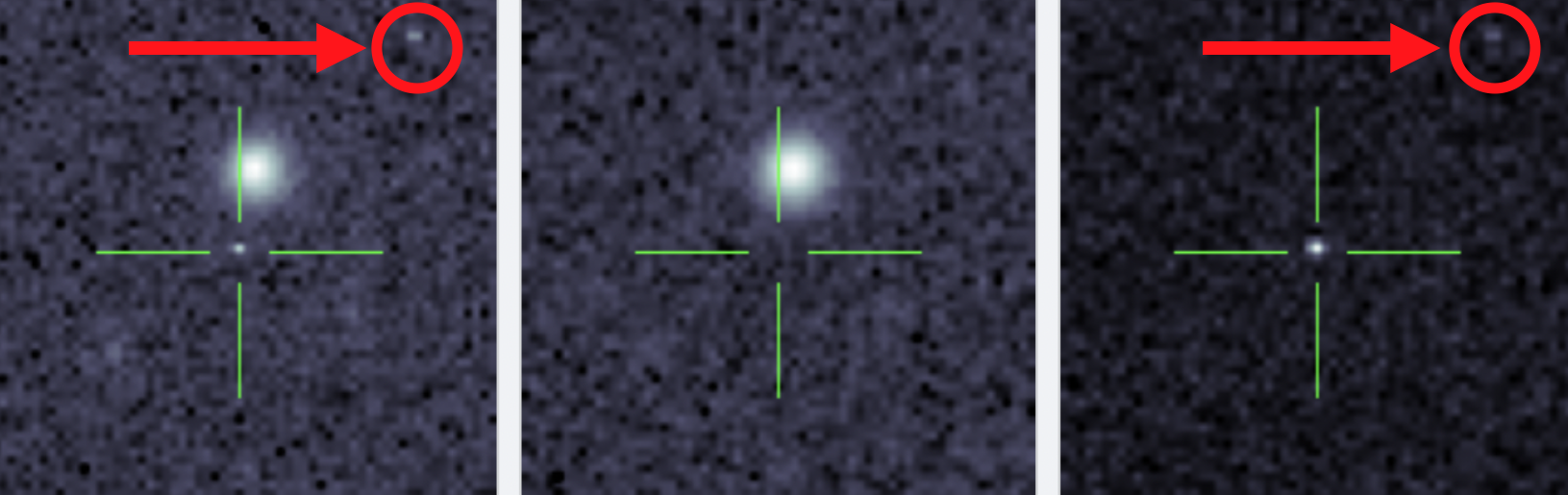}
    \caption{ZTF24aaqqtht can be seen in the science image (left) marked by the crosshair along with a source in the red circle. Both are not present in the reference image (middle); however, both show up in the difference image (right) indicating that depth of the image changed rather than the individual brightness of ZTF24aaqqtht.}
    \label{fig:ZTF24aaqqtht}
\end{figure}

This source only had two detections associated with it, so a period was not able to be found; additionally no \texttt{SCoPe} classifications were computed. ZTF24aaqqtht had a magnitude of 20.5 in the  $g$-band during the time of its detection, which is very close to the detection limit of ZTF \citep{Graham_2019}. This source is likely not a transient because an additional faint source also shows up in the image subtraction (see Fig.~\ref{fig:ZTF24aaqqtht}). A likely explanation is that the source is a cosmic ray since the point spread function is atypical compared to normal detections. 
\subsection{ZTF24aaqrjdd}
\begin{figure}[!htb]
    \centering
    \includegraphics[width=1\linewidth]{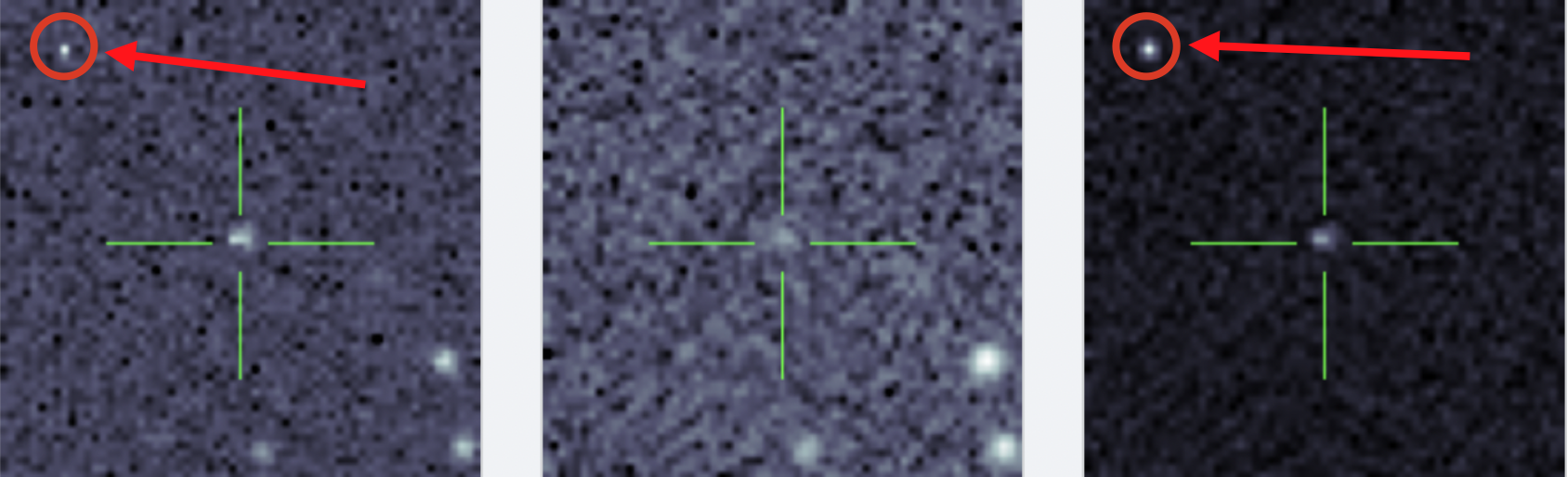}
    \caption{Science image (left), Reference (middle) and Difference image (right) of ZTF24aaqrjdd. Like ZTF24aaqqtht an additional source (red arrow) in the difference and science images indicates some data quality issues.}
    \label{fig:ZTF24aaqrjdd_diff}
\end{figure}

ZTF24aaqrjdd's difference image looks similar to ZTF24aaqqtht (Fig. \ref{fig:ZTF24aaqrjdd_diff}), but unlike ZTF24aaqqtht there is enough data in DR16 for \texttt{SCoPe} to attempt classification. This source has no classification from any algorithm over 0.4. Additional the scores for variable are 0.00 for XBG and only 0.12 for DNN. In PANSTARRS dr2 \citep{panstarrs} at the location of ZTF24aaqrjdd there is an extended source. Given the machine learning scores, the extended sources in other surveys and an ALLWISE color W1-W2 = 1.139, ZTF24aaqrjdd is consistent with a persistent source (a galaxy) and some issue with the image acquisition for the experiment likely occurred as indicated by the bad pixels in Fig. \ref{fig:ZTF24aaqrjdd_diff}.

\subsection{ZTF18aajtlgu}
\texttt{SCoPe} classified ZTF18aajtlgu as an RR Lyrae; there was no strong classification for which RR Lyrae type. The standard pipeline found a period of 1.533\,days. When using only the experiment data a period of 0.255\,days is found. The true period is 0.51135 which is too close to half a day so it is excluded from \texttt{SCoPe}'s period search. ZTF18aajtlgu was also identified by \cite{Sesar_2017} as a RR Lyrae star.
\begin{figure*}[!htb]
    \centering
    \includegraphics[width=1\linewidth]{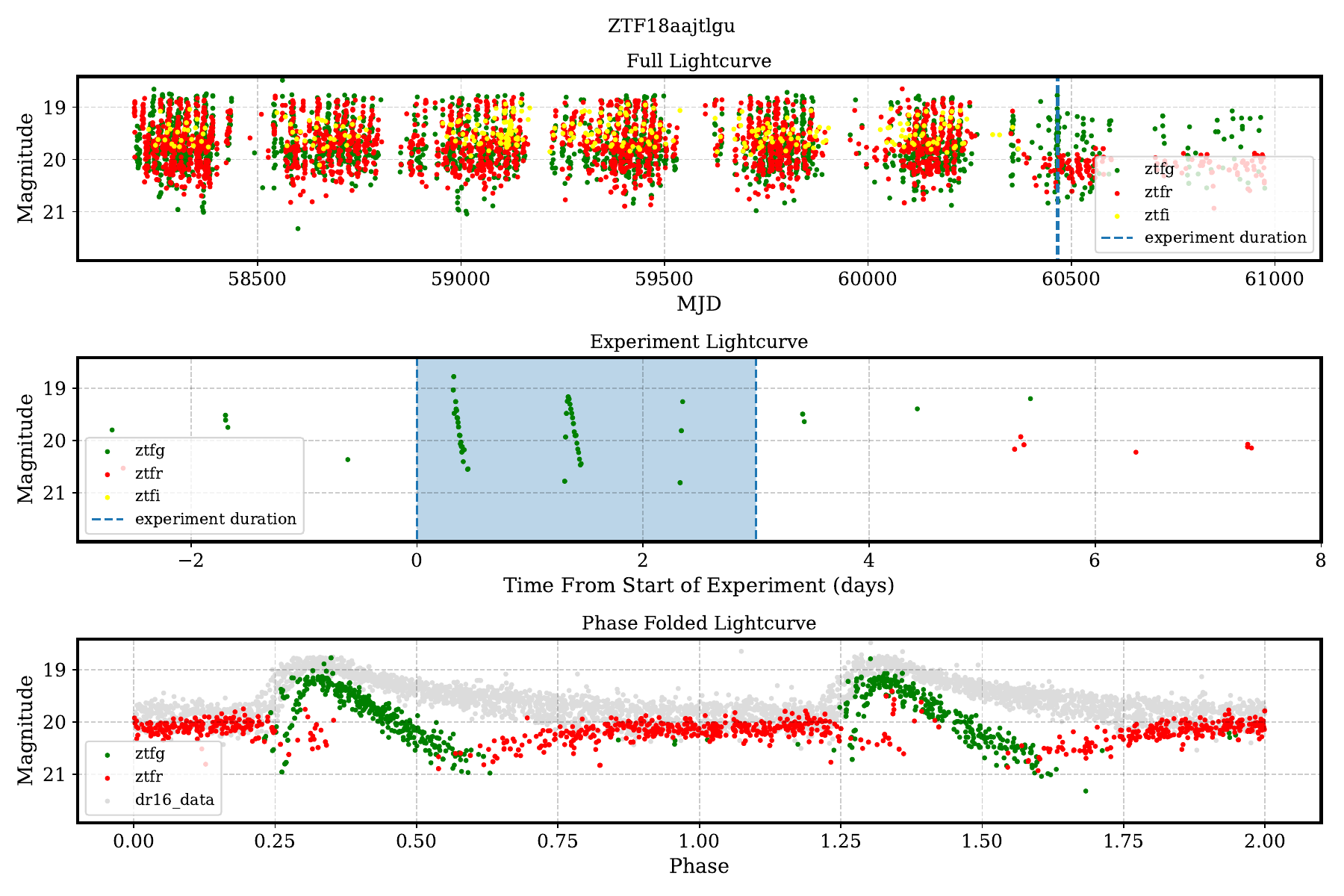}
    \caption{The full DR16 and alert light curve for ZTF18aajtlgu is shown in the top panel. In the shaded region in the middle panel the experiment data is shown and in the bottom panel the light curve is folded at its period of 0.255 days (bottom). Note that ZTF18aajtlgu is not detectable in the ZTFg band during the minima of its period (phase $\approx 0.8$) in the alert data.}
    \label{fig:ZTF18aajtlgu}
\end{figure*}

\subsection{ZTF18aajtkma}
\begin{figure*}[!htb]
    \centering
    \includegraphics[width=1\linewidth]{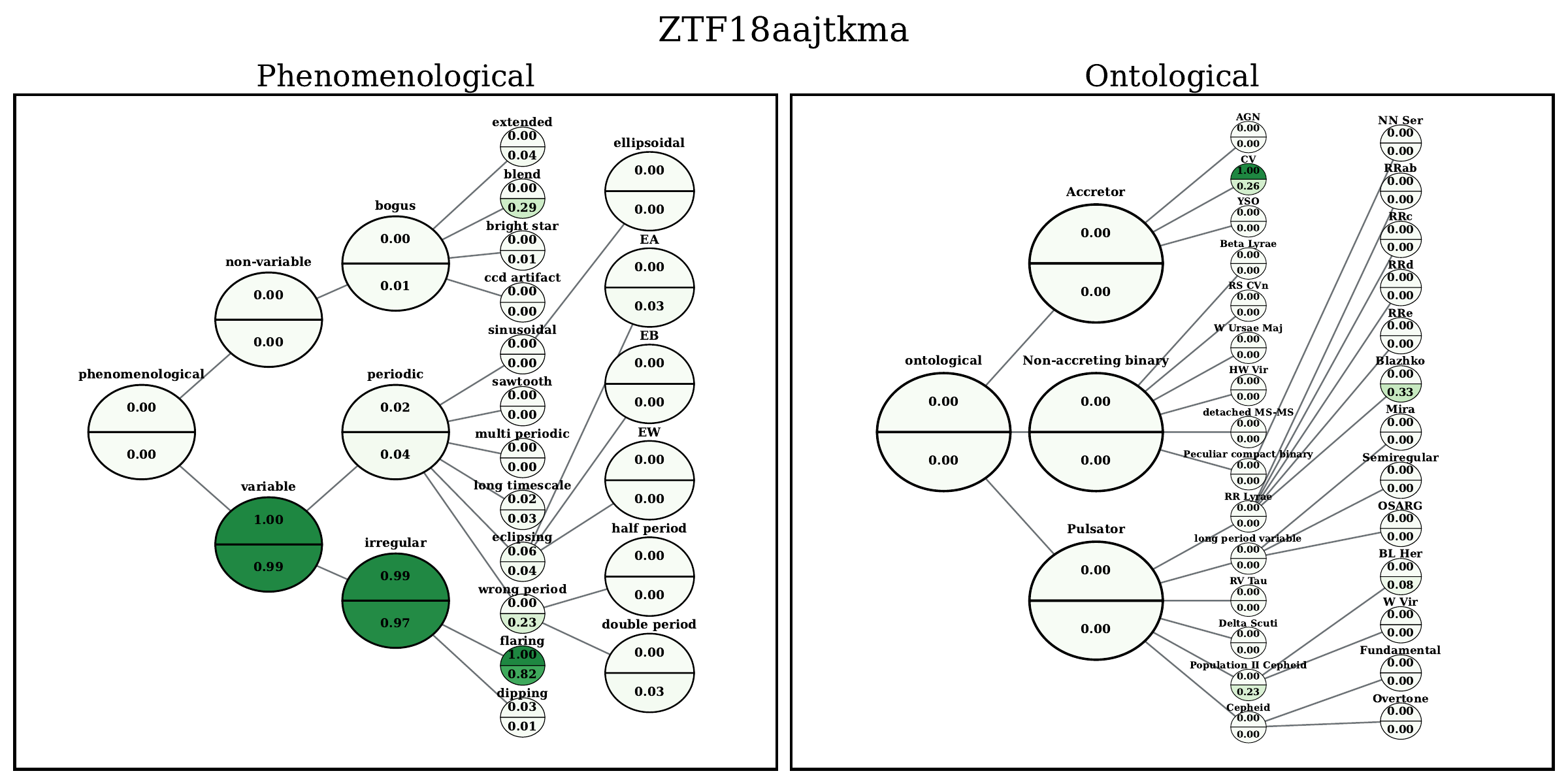}
    \caption{All classification for ZTF18aajtkma. The top panel shows the phenomenological classifications and the bottom show the ontological classifications. The top half of the circles show the DNN score while the lower half show the XGB scores. Note that while the DNN and XGB models disagree on the CV classification (1.00 vs 0.26), both have great agreement on the iregular classification (0.99 vs 0.97).}
    \label{fig:ZTF18aajtkmaClass}
\end{figure*}

\begin{figure*}[!htb]
    \centering
    \includegraphics[width=1\linewidth]{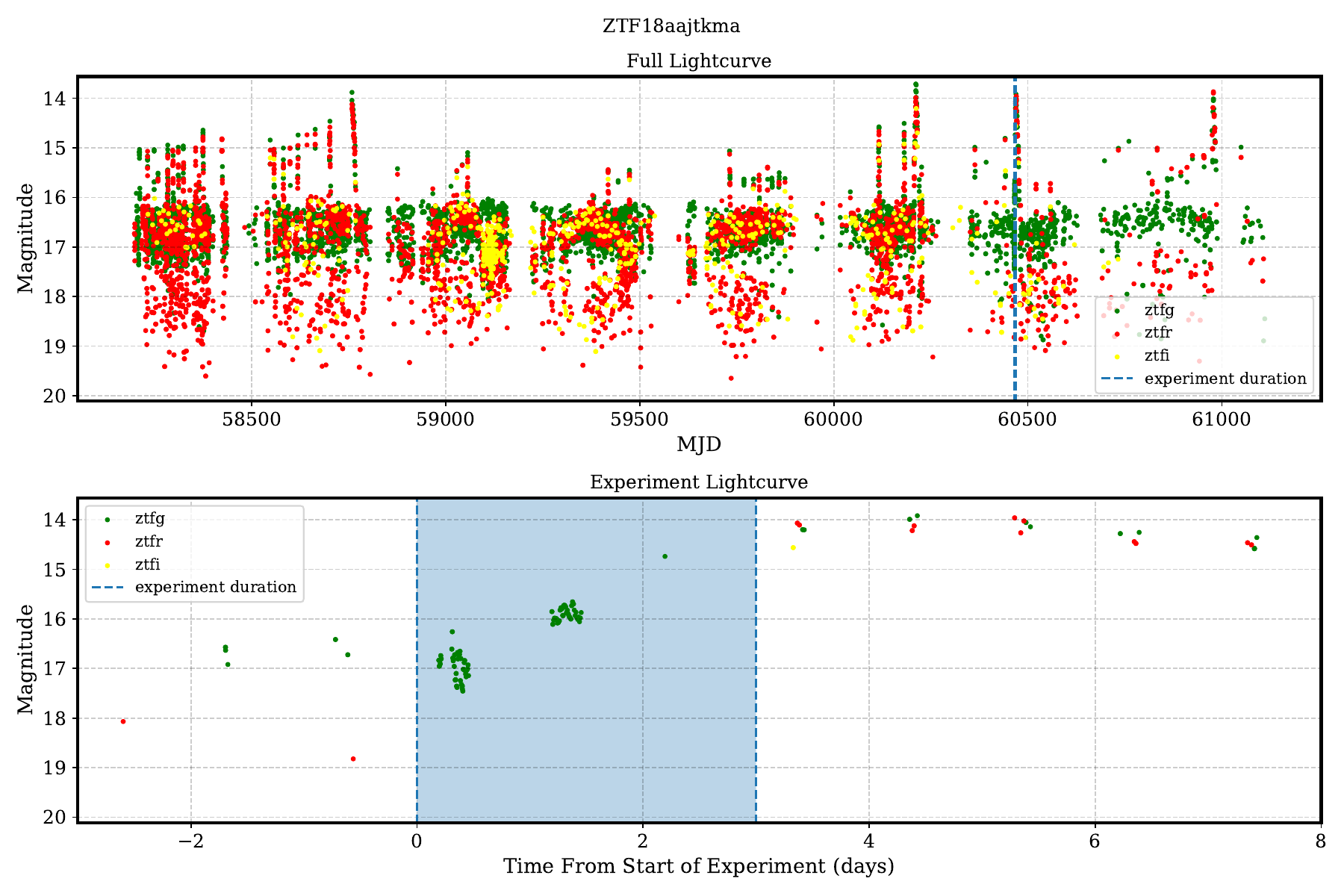}
    \caption{Full light curve of ZTF18aajtkma (top). The light curve just during the experiment (bottom). In the top panel it is clear that a large spike occurred during the time of the experiment.}
    \label{fig:ZTF18aajtkma}
\end{figure*}

ZTF18aajtkma is a known CV, as noted in\cite{aajkma_0riginal}. \texttt{SCoPe}'s DNN model classify it as a CV but the XGB model only note a low classification score ($<.3$); however, both XGB and DNN models show high confidence in the flaring and irregular classification indicating that these classifications may be more robust (Fig. \ref{fig:ZTF18aajtkmaClass}). The flaring behavior of this CV dominated the period finding procedure, causing no relevant period to be found in the experiment data or the full \texttt{SCoPe} pipeline. As seen in Fig.~\ref{fig:ZTF18aajtkma}, a flare occurred during the experiment causing the filter to be triggered.

\subsection{ZTF18aakfqxu}
\begin{figure*}[!htb]
    \centering
    \includegraphics[width=1\linewidth]{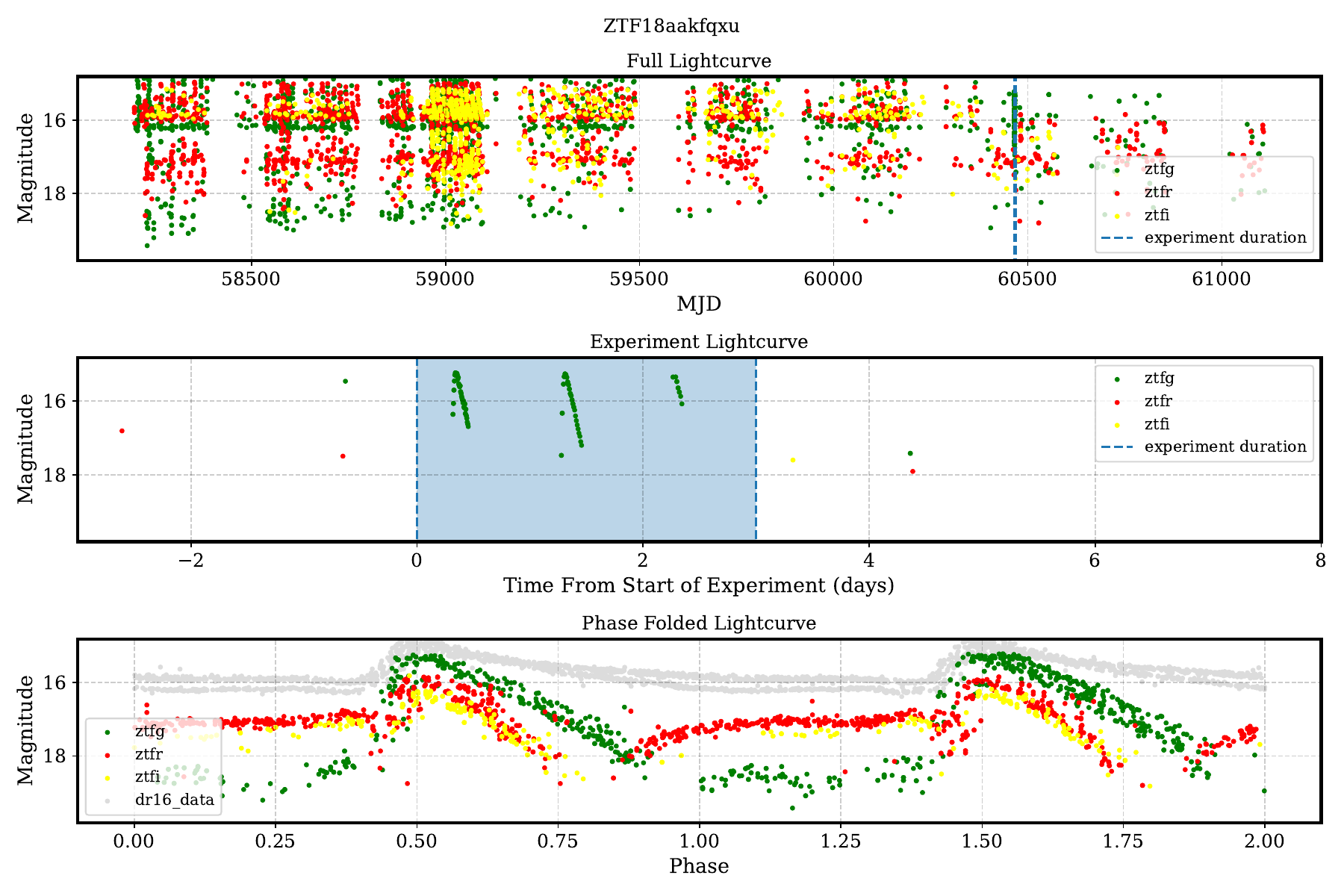}
    \caption{Full light curve of ZTF18aakfqxu in the 3 ZTF bands (top). The light curve just during the experiment (middle) and the light curve folded at its period of 0.241 days (bottom).}
    \label{fig:ZTF18aakfqxu}
\end{figure*}

In the full \texttt{SCoPe} pipeline, this source was classified as an RRab by both DNN and XGB. The period from the full \texttt{SCoPe} pipeline and from only the experiment data was 0.483\,days which is the 2 $\times$ harmonic of the true period. This star was previously classified as an RR Lyrae under the name 2MASS J15045038+6616368 in \cite{Sesar_2013}. Note that in the bottom panel of Fig.~\ref{fig:ZTF18aakfqxu} the light curve varies by more than one magnitude. Since the period is less than one day criteria ii) of the filter was triggered.
\subsection{ZTF18aapnpxp}
\texttt{SCoPe} classified this as an RR Lyrae, non-specific type. \texttt{SCoPe} found a period of 1.471\,days which is a 9 times 0.163\,days which was found using only ZTF-ULTRASAT experiment data. The true period of ZTF18aapnpxp is 0.490 days which is excluded from \texttt{SCoPe}'s search as it is too close to half a day. This source was also identified as an RR Lyrae by \cite{aapnpxp_original}.
\begin{figure*}[!htb]
    \centering
    \includegraphics[width=1\linewidth]{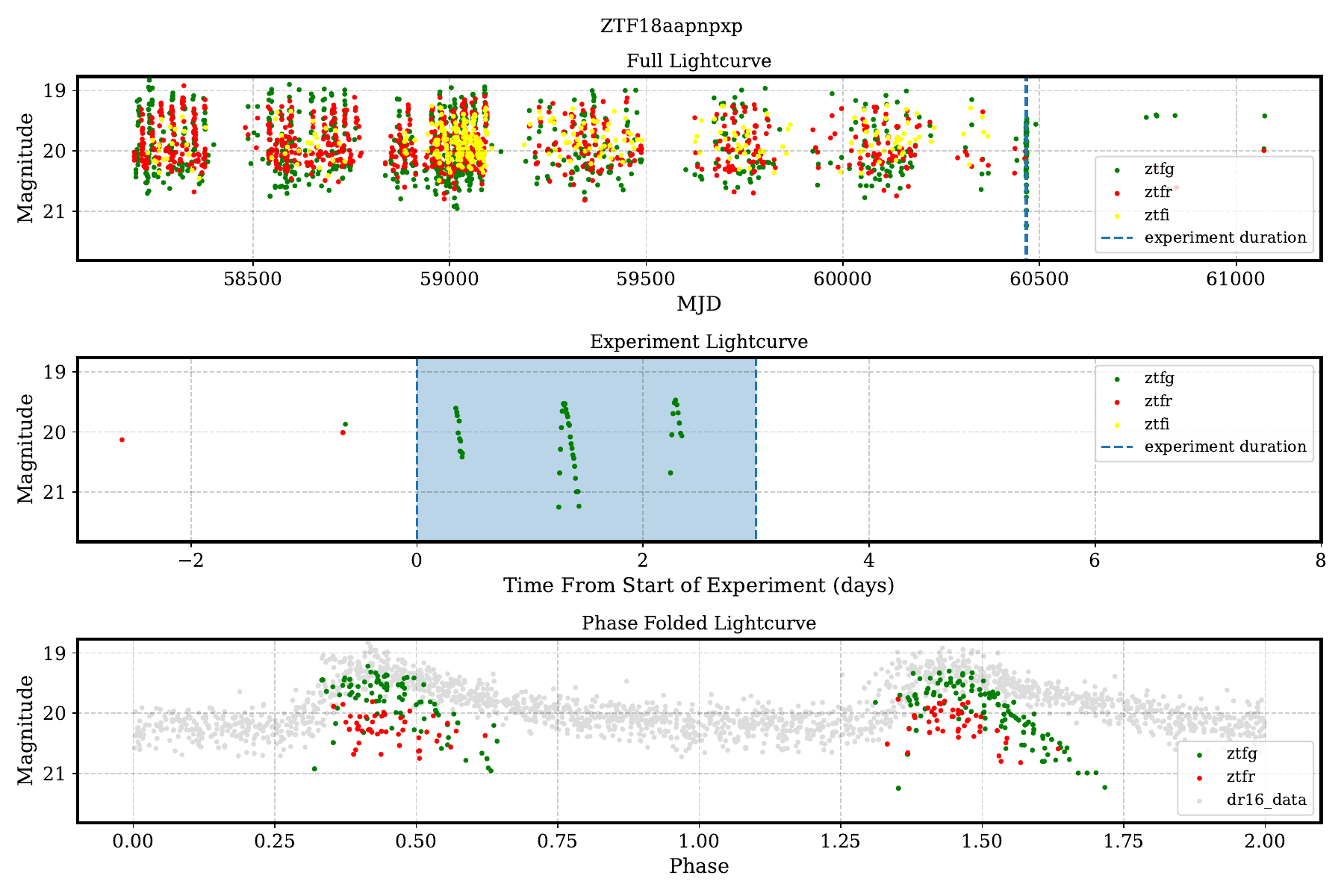}
    \caption{Full light curve of ZTF18aapnpxp in the 3 ZTF bands (top). The light curve just during the experiment (middle) and the light curve folded at its period of 0.490 days (bottom).}
    \label{fig:ZTF18aapnpxp}
\end{figure*}
\subsection{ZTF21aasjkbd}
\begin{figure*}[!htb]
    \centering
    \includegraphics[width=1\linewidth]{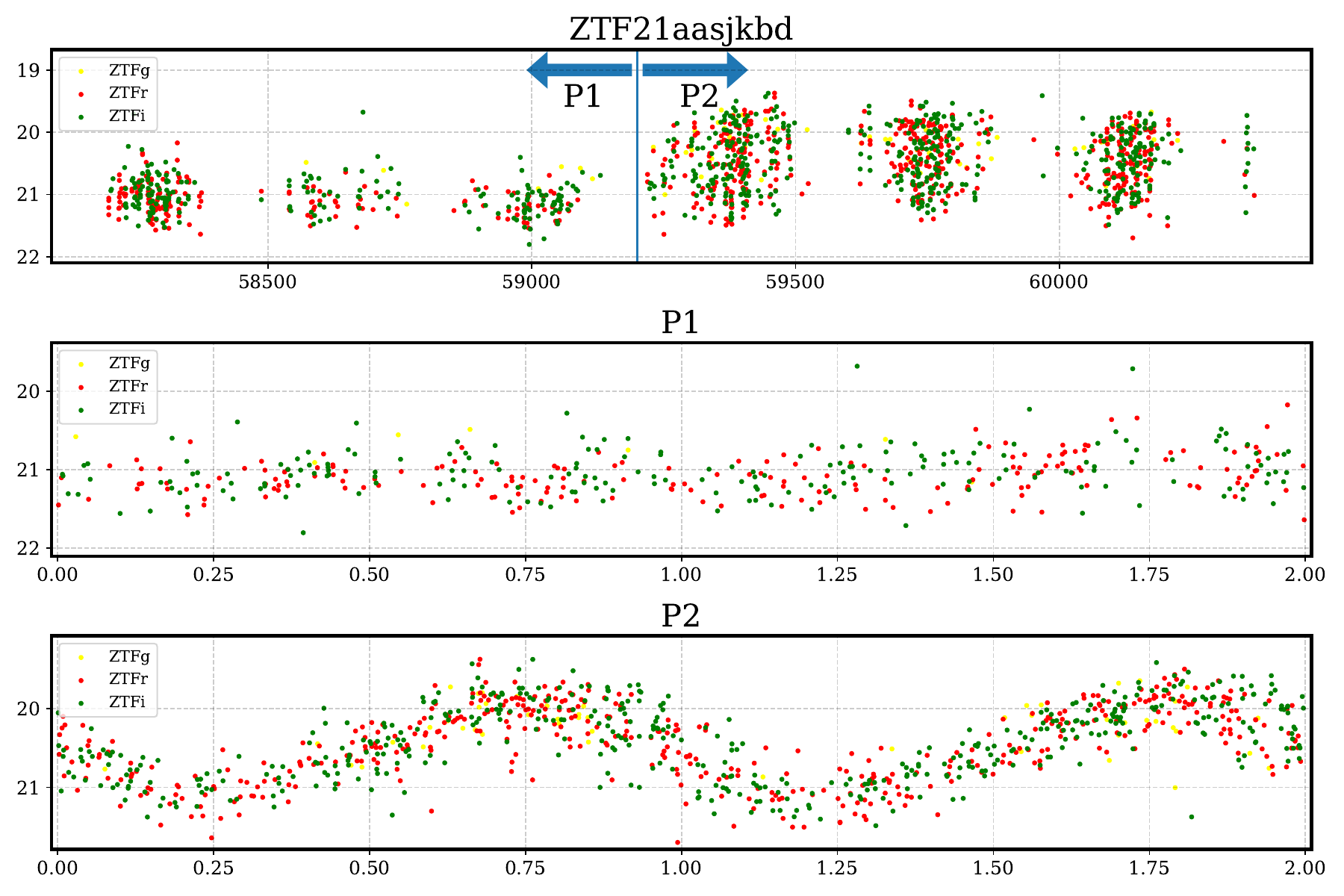}
    \caption{ZTF21aajkbd exhibits two phases in its history. In P1 (middle) there is no periodic variability at \texttt{SCoPe}'s found 1.216 days, but in P2 (bottom) the periodic variability is clear.}
    \label{fig:ZTF21aasjkbd_lc}
\end{figure*}

Upon first inspection ZTF21aasjkbd appears to be a AGN due to the variability seen in its optical lightcurve (see Fig. \ref{fig:ZTF21aasjkbd_lc}) and is source was classified as a QSO in PS1-STRM and has a W1-W2= 0.7 in the CATWISE \citep{CATWISE} catalog which further supports that this source is an AGN. However \texttt{SCoPe} did not classify this source as an AGN instead it was classified as flaring for dnn and irregular for xgb. The full \texttt{SCoPe} pipeline found a period of 1.216 days, but the periodic variability only shows up around December 2020 (mjd 59200, P2 in Fig. \ref{fig:ZTF21aasjkbd_lc}). Before that date (P1) there is no periodic variability at the period determined by \texttt{SCoPe}. The change in 2020 is also accompanied by brightness increase of about half a magnitude. Further investigation is need to determine what phenomena produced this source.

\section{Conclusions}
\label{sec:conclusion}

In this paper, we present the results of a pilot experiment designed to anticipate the challenges ULTRASAT will face in real-time transient detection. We use ZTF observations to simulate high-cadence fields, apply real-time filtering, and evaluate how variable sources and flaring objects may produce false positives. We also demonstrate the utility of machine-learning-based catalogs, specifically \texttt{SCoPe}, for automated classification and filtering.

During the experiment, the primary types of candidates to pass the filter were RR Lyrae stars and a Cataclysmic Variable (CV). RR Lyrae are short-period variables with high amplitudes. Many of these RR Lyrae are near ZTF's magnitude limit, so during their minima they are undetectable, causing them to appear as “new” sources during their pulsations. Such sources can be effectively excluded by anti-matching the SCoPe catalog for amplitudes greater than 0.8\,mag and periods shorter than one day. For the ZTF-ULTRASAT experiment covering fields 848, 847, and 825, this corresponds to 1,018 light curves with periods (significance $>$ 10) less than a day and amplitudes greater than 0.8\,mag, out of roughly 4.5 million light curves.

CVs, while also periodic, do not regularly trigger the experimental filter based on period or amplitude. Instead, alerts are generated by their flaring behavior, which is captured by the \texttt{SCoPe} classifications fla\_dnn and fla\_xgb. These flaring scores range from 0 to 1, allowing flexibility in threshold selection for exclusion. Fig.~\ref{fig:SkyThresh} shows how the density of sources varies as a function of classification threshold.

Over the three nights of the experiment, only seven alerts passed the applied filter. As expected, flaring sources such as ZTF18aajtkma triggered the filter, but it was somewhat surprising that RR Lyrae also passed. Short-period, high-amplitude variables can mimic transient signatures when observed over limited time windows, highlighting the need for a detailed catalog of such sources in future surveys. The \texttt{SCoPe} catalog provides periods, amplitudes, and machine-learning classifications for flaring objects, enabling automatic cross-matching to exclude these sources in transient searches. In this study, the primary false positives were RR Lyrae and flaring sources, representing only 4 of the 88 \texttt{SCoPe} classification categories. In future work, the remaining 84 categories could be leveraged to construct more robust filters or to exclude additional classes of false positives not encountered in this experiment.

\begin{figure}[!htb]
    \centering
    \includegraphics[width=1\linewidth]{{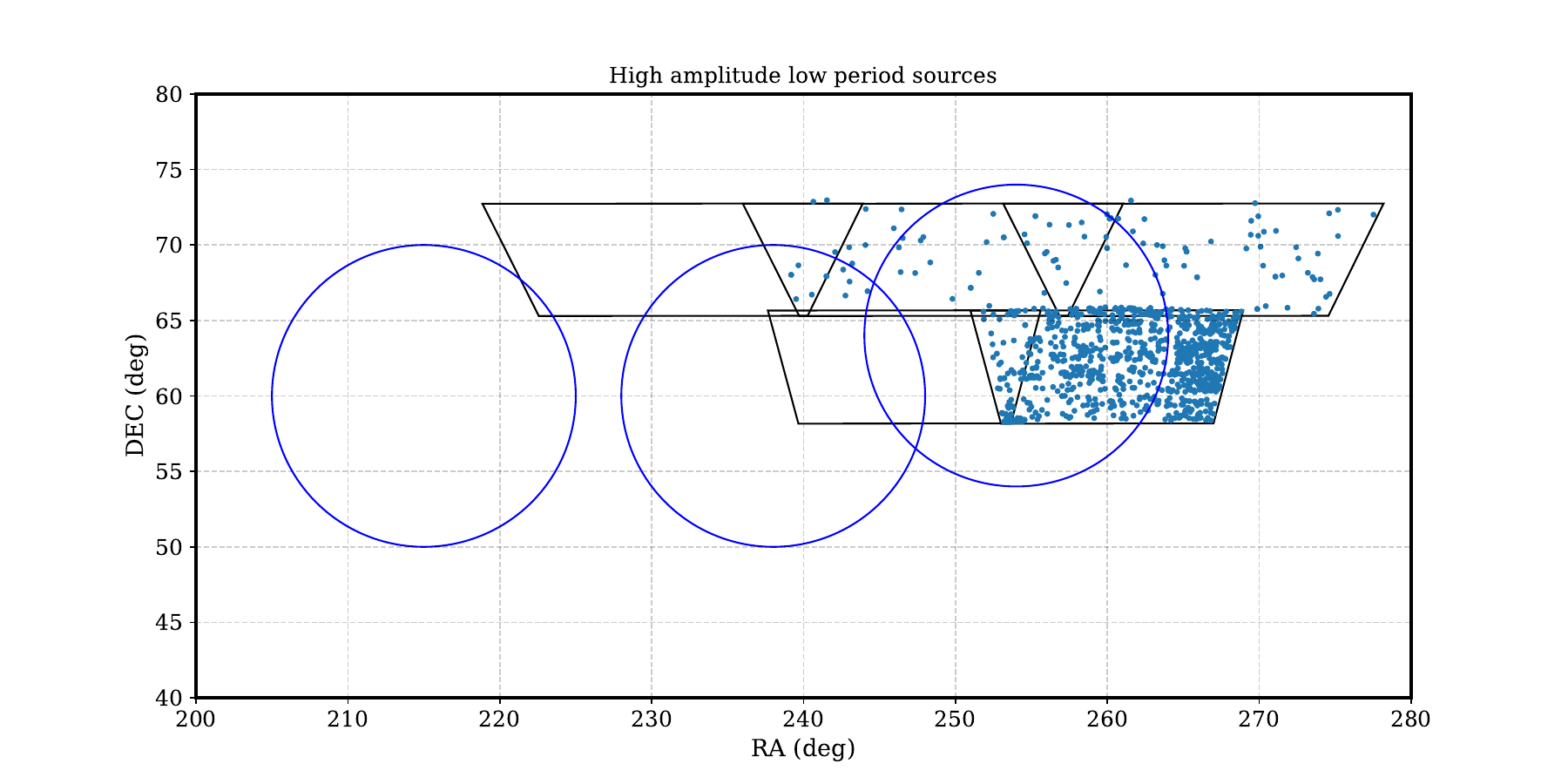}}
    \caption{Circles denote the ULTRASAT northern high cadence field, Black outlined regions are the ZTF fields from the experiment, blue points are light curves with periods less than a day and amplitudes greater than 0.8 which would trigger the real time filter used for the experiment.}
    \label{fig:HighAmp_LowPer_ULTRA}
\end{figure}

Looking forward, ULTRASAT will encounter a substantially larger volume of alerts than was tested in this experiment. The combination of its wide field of view, high-cadence coverage, and UV sensitivity will produce numerous transient candidates, including both rare astrophysical events and common variable sources. Leveraging catalogs such as \texttt{SCoPe} will be critical for automated filtering, allowing high-confidence identification of genuine transients while excluding known variable stars and flaring sources. In addition, \texttt{SCoPe}’s full set of 88 classifications provides the potential to preemptively identify other classes of false positives, including eclipsing binaries, long-period variables, or AGN, further improving the purity of transient alerts.

By cross-matching incoming ULTRASAT alerts with a well-characterized variable star catalog and applying classification-based thresholds, future surveys will be able to focus follow-up resources on high-value targets such as kilonovae, early supernova shock breakouts, and other short-lived phenomena. The approach demonstrated here—combining real-time filtering with machine-learning-based catalogs—establishes a scalable framework that will be essential for managing the high alert rates expected from ULTRASAT, ensuring efficient and accurate transient discovery across its wide UV sky coverage.

\begin{figure}[!htb]
    \centering
    \includegraphics[width=1\linewidth]{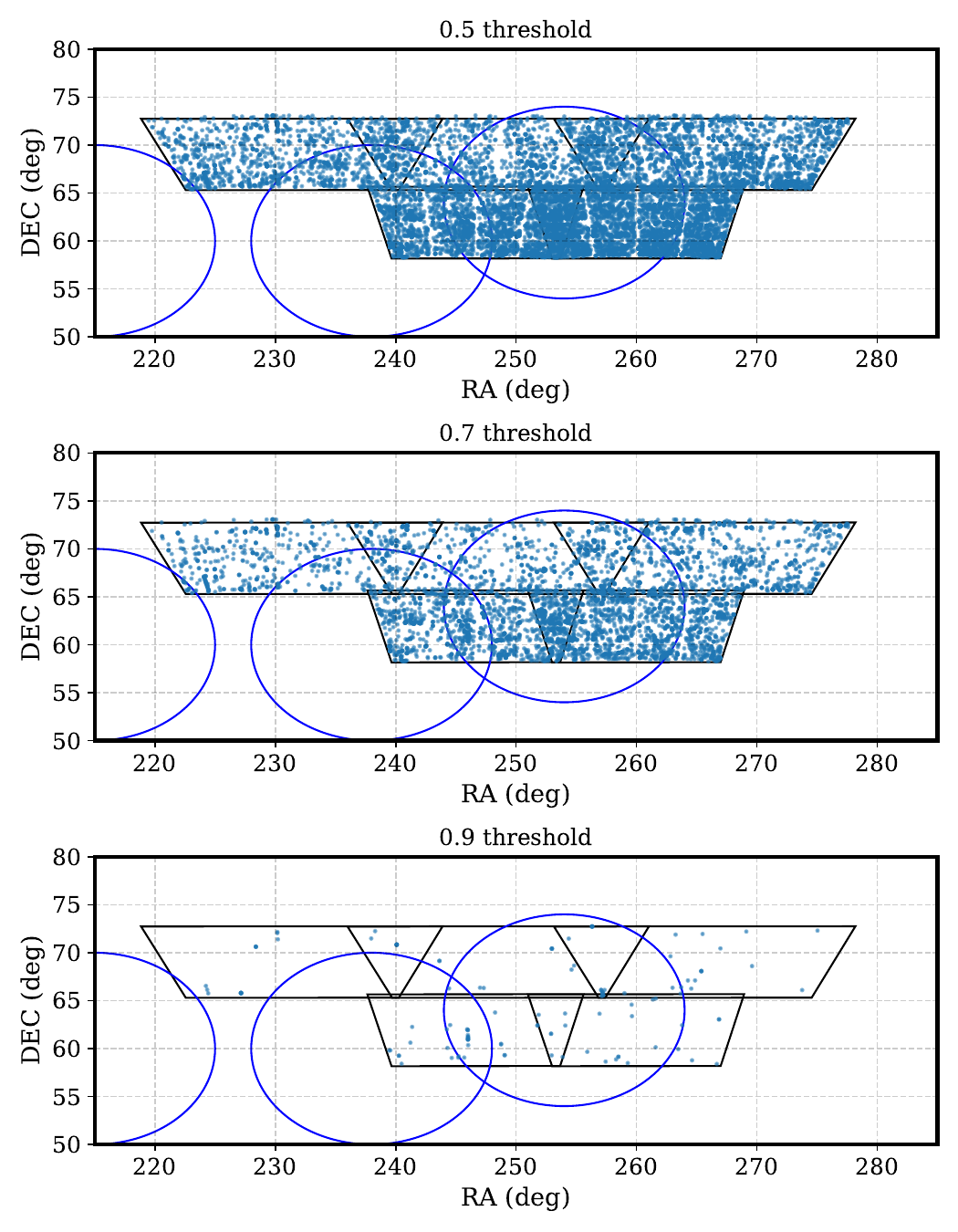}
    \caption{The number and density of light curves classified as flaring changes as the threshold for classification becomes more strict from top to bottom.}
    \label{fig:SkyThresh}
\end{figure}

\section{acknowledgments}
\label{sec:acknowledgments}

Based on observations obtained with the Samuel Oschin Telescope 48-inch and the 60-inch Telescope at the Palomar Observatory as part of the Zwicky Transient Facility project. ZTF is supported by the National Science Foundation under Grants No. AST-1440341, AST-2034437, and currently Award No. 2407588. ZTF receives additional funding from the ZTF partnership. Current members include Caltech, USA; Caltech/IPAC, USA; University of Maryland, USA; University of California, Berkeley, USA; University of Wisconsin at Milwaukee, USA; Cornell University, USA; Drexel University, USA; University of North Carolina at Chapel Hill, USA; Institute of Science and Technology, Austria; National Central University, Taiwan, and OKC, University of Stockholm, Sweden. Operations are conducted by Caltech's Optical Observatory (COO), Caltech/IPAC, and the University of Washington at Seattle, USA.

E.O.O. is grateful for the support of Paul and Tina Gardner,
The Norman E Alexander Family M Foundation ULTRASAT Data Center Fund,
Israel Science Foundation,
Minerva, and Israel Council for Higher Education (VATAT).

The Gordon and Betty Moore Foundation, through both the Data-Driven Investigator Program and a dedicated grant, provided critical funding for SkyPortal.

D.E.W. and M.W.C. acknowledge support from the National Science Foundation with grant numbers PHY-2117997, PHY-2308862 and PHY-2409481.



\bibliographystyle{aasjournalv7}
\bibliography{main} 

\label{lastpage}
\end{document}